\documentclass[
twocolumn]{aastex701}

\usepackage{amsmath, amssymb}

\begin{document}

\title{Constraining the Planetary Obliquity Distribution of Warm Jupiters}

\author[0000-0001-9985-0643]{Caleb Lammers}
\affiliation{Department of Astrophysical Sciences, Princeton University, 4 Ivy Lane, Princeton, NJ 08544, USA}
\email{caleb.lammers@princeton.edu}

\author[0000-0002-4265-047X]{Joshua N.\ Winn}
\affiliation{Department of Astrophysical Sciences, Princeton University, 4 Ivy Lane, Princeton, NJ 08544, USA}
\email{jnwinn@princeton.edu}

\shorttitle{Warm Jupiter Obliquities}
\shortauthors{Lammers \& Winn}

\begin{abstract}

Warm Jupiters are an intriguing class of planets with uncertain origins. Their planetary obliquities could help distinguish between different formation pathways: planet-planet scattering and migration across resonances can excite large obliquities, whereas in-situ formation would more naturally produce low obliquities. We searched for oblateness-related anomalies in the transit light curves of six observationally favorable warm Jupiters: TOI-201\,b, TOI-1670\,c, TOI-199\,b, Kepler-9\,c, Kepler-30\,c, and Kepler-553\,c. Each planet's light curve is consistent with a spherical planet and provides degenerate constraints on the planet's sky-projected oblateness and obliquity. To overcome these limitations, we performed hierarchical Bayesian modeling of the population-level obliquity distribution. Assuming warm Jupiters are as oblate as Saturn ($f\,{\approx}\,0.1$), we find their median obliquity to be below $12^\circ$ with $90$\% confidence and below Saturn's obliquity ($27^\circ$) with $93$\% confidence. Jupiter-like oblateness ($f\,{\approx}\,0.06$) and larger obliquities are allowed. Simulations of {\it JWST} observations predict that significantly tighter constraints can be derived.

\end{abstract}

\keywords{\uat{exoplanets}{498} --- \uat{oblateness}{1143} --- \uat{transit photometry}{1709}}

\section{Introduction}
\label{sec:intro}

Warm Jupiters occupy orbits large enough to prevent tidal evolution from erasing clues to their origin, yet small enough for transit and Doppler surveys to find them. Thus, neither too hot nor too cold, they are just right for studying giant-planet formation and migration. Warm Jupiters are typically defined as giant planets with periods between about $10$~days and a year, and occur around a few percent of Sun-like stars \citep[e.g.,][]{Wittenmyer2020}. Warm Jupiters differ from hot Jupiters in at least three ways: their orbital eccentricities span a broader range \citep{Dong2014, Dong2021}, presumably because tidal circularization is slow or negligible; their host stars' obliquities tend to be lower \citep{Rice2022, Wang2024}; and they are more likely to have nearby companions \citep{Steffen2012, Huang2016, Wu2023}.

Despite these hints, the formation of warm Jupiters remains uncertain. The existence of nearby companions has led some authors to advocate for in-situ formation \citep{Huang2016, Boley2016}. The broad eccentricity distribution has led others to argue for planet-planet scattering, possibly following in-situ formation or disk-driven migration \citep{Mustill2017, Anderson2020, Dong2026}. Eccentric warm Jupiters could also be produced by secular dynamical interactions with exterior companions \citep{PetrovichTremaine2016, AndersonLai2017}. Additional diagnostics would be helpful, especially if they probe dynamical histories from a different angle.

Measuring the spin-axis orientations of warm Jupiters would suit this purpose. In the absence of dynamical interactions, planets are expected to form with low obliquities. Migration into resonant configurations \citep{Millholland2024, Lu2025} and collisions with other planets \citep{Li&Lai2020, Li&Lai2021} might excite planetary obliquities to values as large as $90^\circ$. Thus, a broad range of planetary obliquities would support the migration and planet-planet scattering formation scenarios, while
predominantly low obliquities would be compatible with in-situ formation.

Planetary obliquities are challenging to measure. Constraints have been reported for a few planets by comparing orbital inclinations (from astrometry) with planetary spin-axis inclinations (from radii, rotation periods, and projected rotation speeds; \citealt{Bryan2020, Bryan2021, PalmaBifani2023, Poon2024, Gandhi2025}). Although groundbreaking, such analyses are limited to directly imaged companions, which typically have masses ${\gtrsim}\,10\,\mathrm{M_{Jup}}$ and orbital separations ${\gtrsim}\,50$\,AU.

For warm Jupiters, precise transit photometry provides another possible route to obliquities. The light curve of an oblate planet differs slightly from a spherical planet of the same projected area \citep{Seager&Hui2002, Barnes&Fortney2003}. However, even for giant planets, the photometric deviations are ${\sim}\,100$~parts per million (ppm) and occur primarily during the brief ingress and egress phases, accentuating the challenge. Nonetheless, progress has been made. \citet{Carter&Winn2010} derived the first empirical constraints on an exoplanet's oblateness, for the hot Jupiter HD\,189733\,b. Their upper limits on oblateness were stringent but unsurprising, because tidal despinning and spin-orbit alignment are expected to drive the oblateness and obliquity of a hot Jupiter beneath detectability limits. Investigators have since applied this technique using {\it Kepler} data \citep{Zhu2014, Price2025} and {\it JWST} data \citep{LammersWinn2024, Liu2024, Dholakia2025, Cassese2026}, in all cases reporting upper limits for individual targets.

For an individual system, the interpretation of an upper limit is complex because of parameter degeneracies and sky-projection effects. Even if the planet is oblate, a non-detection can occur due to a low obliquity or because its sky projection happens to be nearly circular, rendering the signal imperceptible. We were thereby motivated to develop a hierarchical Bayesian method to overcome these limitations, and to identify a sample of warm Jupiters for which available data could lead to meaningful constraints. Our search led us to six planets for which suitable data were obtained by the {\it Kepler} mission \citep{Borucki2010} and the {\it TESS} mission \citep{Ricker2015}.

\section{Sample selection}
\label{sec:targets}

Identifying favorable oblateness targets involves several important considerations. Obviously, transit observations with a very high signal-to-noise ratio (SNR) are a necessity. Less obviously, the transit impact parameter $b$ is critical. The oblateness-induced perturbation to a light curve can be decomposed into functions that have even or odd symmetry with respect to the transit midpoint. Although these components generally have similar amplitudes, the even component is much more difficult to detect due to strong degeneracies with other transit parameters, as illustrated in Figure~\ref{fig:antisymmetric}. The odd component's amplitude is proportional to $b\sqrt{1\,{-}\,b^2}$, as derived in Appendix~\ref{sec:oblate_geometry}. Thus, the impact parameter strongly affects signal detectability, and the optimal value is $b\,{=}\,2^{-1/2}\,{\approx}\,0.7$ \citep{Barnes&Fortney2003}.

To avoid the problem of tidal despinning, the planet should be
sufficiently far from the star. The approximate timescale for spin-orbit synchronization based on equilibrium-tide theory is \citep{Goldreich&Soter1966, Carter&Winn2010b}
\begin{align}
\label{eqn:tau_synch}
\tau_s \approx 1.22\,\mathrm{Gyr} &\times \left(\frac{Q_p'}{10^6}\right) \left(\frac{\mathcal{C}}{0.25}\right) \left(\frac{10\,\mathrm{hr}}{P_\mathrm{rot}} \right) \nonumber\\
&\times \left(\frac{M_p}{\mathrm{M_{Jup}}} \right) \left(\frac{\mathrm{R_{Jup}}}{R_p} \right)^{\!\!3} \left(\frac{P_\mathrm{orb}}{20\,\mathrm{days}} \right)^{\!\!4},
\end{align}
where $Q_p'\,{=}\,3Q_p/(2k_2)$ is the planet's modified tidal quality factor, $P_\mathrm{rot}$ is its rotation period, and $\mathcal{C}$ is its moment of inertia divided by $M_p\,R_p^2$. For the Solar System's giants, interior models suggest $\mathcal{C}\,{\approx}\,0.25$ \citep{Hubbard&Marley1989, Militzer&Hubbard2023}, and $Q_p'$ is estimated to be in the range $10^5$\,--\,$10^{6.5}$ \citep{Goldreich&Soter1966, Ogilvie&Lin2004, Jackson2008}. We adopted the nominal values of $\mathcal{C}\,{=}\,0.25$, $Q_p'\,{=}\,10^6$, and $P_\mathrm{rot}\,{=}\,10$\,hr.

\begin{figure*}
\centering
\includegraphics[width=0.95\textwidth]{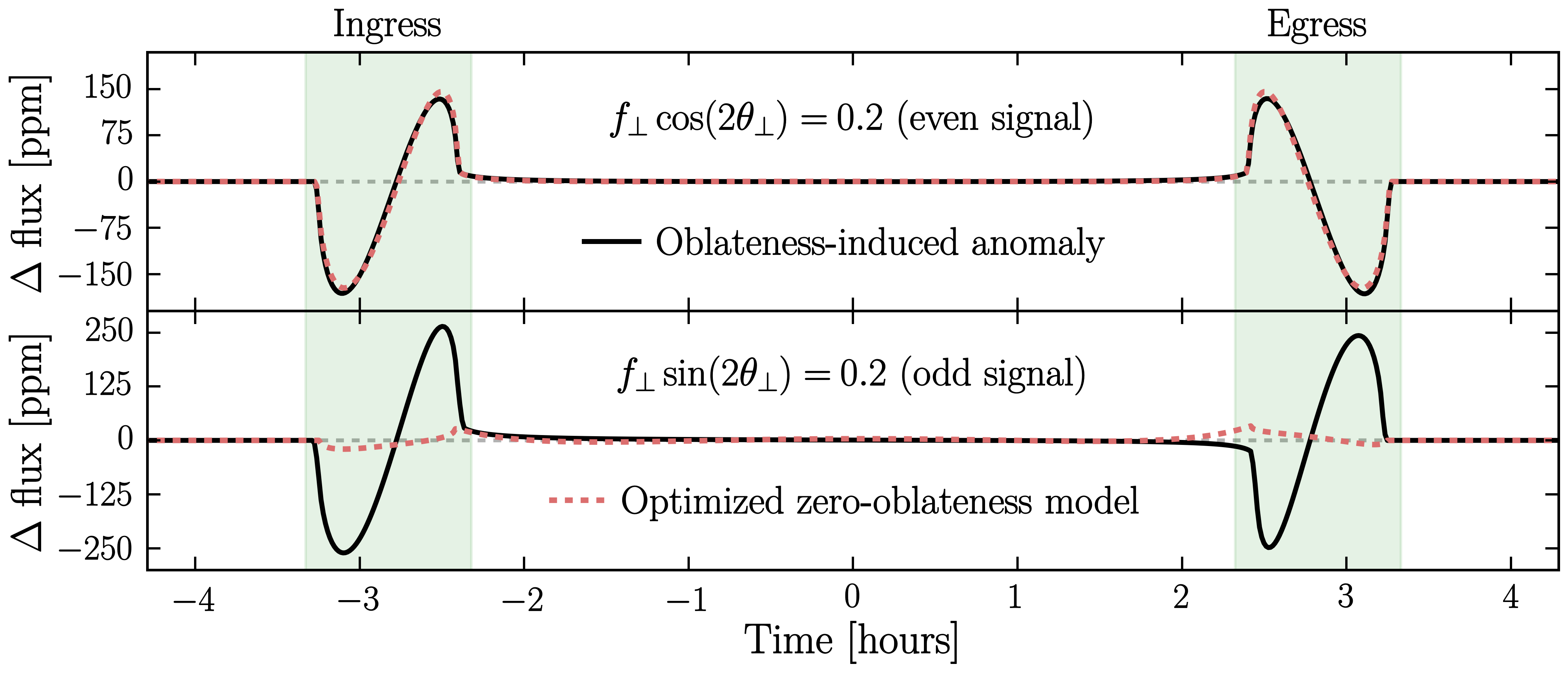}
\caption{The even-symmetry component of the oblateness signal (top panel) is much harder to detect than the odd-symmetry signal (bottom panel). The black curves show the deviations between an oblate-planet model and the spherical-planet model for TOI-199\,b, obtained by setting $f_\perp\,{=}\,0$ without optimizing the other transit parameters ($f_\perp$ and $\theta_\perp$ are defined in Section~\ref{sec:oblate_model}). Green regions are ingress and egress, where the deviations are largest. Dashed red curves show the deviations after optimizing the parameters of the spherical-planet model. Although the even component can be mimicked almost exactly, the odd component still stands out.}
\label{fig:antisymmetric}
\end{figure*}

With these considerations in mind, we began by collecting the relevant parameters of all confirmed transiting planets from the NASA Exoplanet Archive\footnote{\url{exoplanetarchive.ipac.caltech.edu}, accessed June 2026.} \citep{Christiansen2025} and calculating an approximate stacked-transit SNR metric:
\begin{equation}
\label{eqn:transit_SNR}
\mathrm{{SNR}} \approx \frac{\delta}{\sigma_0} \sqrt{N} \propto \left(\frac{R_p}{R_\star}\right)^{\!\!2} \left(\frac{T_\mathrm{dur}\,N_\mathrm{tra}}{10^{0.4V}}\right)^{\!0.5}~.
\end{equation}
Here, $\delta$ is the transit depth, $\sigma_0$ is the photometric precision assuming photon-limited measurements, $N$ is the total number of data points across all transits, and $V$ is
the apparent visual magnitude.

To reduce the risk of tidal despinning, we required $\tau_s\,{\gtrsim}\,1$\,Gyr. The surviving systems were sorted according to SNR\,$\times\,b\sqrt{1-b^2}$. After removing some otherwise promising planets for which the existing data are not suitable\footnote{For Kepler-86\,b and Kepler-1514\,b, short-cadence data are not available. For KOI-12\,b and Kepler-396\,c, complex stellar variability interferes with light-curve modeling.}, the six most favorable cases consisted of three {\it Kepler} systems and three {\it TESS} systems. Table~\ref{table:planet_properties} provides their key properties. The planets range in radius from $0.7$ to $1.1$~R$_\mathrm{Jup}$, in mass from $0.1$ to $7$~M$_\mathrm{Jup}$, and in orbital period from $39$ to $328$ days. Their orbital eccentricities range from $0.0$ to $0.3$, which suggests varying levels of dynamical excitation, although each target resides in a multiplanet system. The hosts are main-sequence stars ranging in mass from $0.9$ to $1.3$~M$_\odot$. Because our selection function was driven mainly by SNR and $b$, which are observer-dependent and unrelated to planetary spin parameters, we expect the sample's oblateness-obliquity distribution to be fairly representative of warm Jupiters around solar-type stars.

\section{Transit photometry}
\label{sec:observations}

We downloaded the {\it Kepler} and {\it TESS} light curves for our six targets from the Mikulski Archive for Space Telescopes using the \texttt{lightkurve} package \citep{Lightkurve2018}. For {\it TESS}, we selected the Science Processing Operations Center (SPOC) light curves \citep{Jenkins2016}. Whenever possible, we used short-cadence data ($1$-min for {\it Kepler} and $20$-sec for {\it TESS}); otherwise, we used long-cadence data ($30$-min for {\it Kepler} and $2$-min for {\it TESS}). When fitting {\it Kepler} long-cadence data, we super-sampled the transit light-curve model at $10$ equally spaced times and calculated the average flux before comparing with the data.

\begin{deluxetable*}{cccccccccccc}
\tablecaption{Summary of Target Properties\label{table:planet_properties}}
\tablehead{
  \colhead{Name\tablenotemark{1}} & \colhead{$R_\star$} & \colhead{$M_\star$} &
  \colhead{$V$} & \colhead{$P_\mathrm{orb}$} & \colhead{$R_p$} &
  \colhead{$M_p$} & \colhead{$e$} & \colhead{$T_\mathrm{dur}$} &
  \colhead{$b$} & \colhead{$N_\mathrm{tra}$} &
  \colhead{$\tau_\mathrm{s}$ (Eq.~\ref{eqn:tau_synch})} \\
  \colhead{} & \colhead{[R$_\odot$]} & \colhead{[M$_\odot$]} &
  \colhead{[mag]} & \colhead{[days]} & \colhead{$[\mathrm{R_{Jup}}]$} &
  \colhead{$[\mathrm{M_{Jup}}]$} & \colhead{} & \colhead{[hours]} &
  \colhead{} & \colhead{} & \colhead{[Gyr]}
}
\startdata
TOI-201\,b & $1.3$ & $1.3$ & $9.1$ & $53$ & $1.1$ & $0.6$ & $0.3$ & $4.8$ & $0.73$ & $18$ & $28$ \\
TOI-1670\,c & $1.3$ & $1.2$ & $9.9$ & $41$ & $1.0$ & $0.6$ & $0.1$ & $5.4$ & $0.76$ & $20$ & $14$ \\
TOI-199\,b & $0.8$ & $0.9$ & $10.7$ & $105$ & $0.8$ & $0.2$ & $0.1$ & $6.7$ & $0.45$ & $9$ & $294$ \\
Kepler-9\,c & $1.0$ & $1.0$ & $13.9$ & $39$ & $0.7$ & $0.1$ & $0.1$ & $4.6$ & $0.74$ & $38$ & $4$ \\
Kepler-30\,c & $1.0$ & $1.0$ & $15.7$ & $60$ & $1.1$ & $2.0$ & $0.0$ & $6.7$ & $0.40$ & $23$ & $135$ \\
Kepler-553\,c & $0.9$ & $0.9$ & $15.0$ & $328$ & $1.0$ & $6.7$ & $0.3$ & $12.3$ & $0.75$ & $5$ & $6\,{\times}\,10^5$
\enddata
\tablenotetext{1}{TOI-201\,b: \citet{MaciejewskiLoboda2025}; TOI-1670\,c: \citet{Tran2022}; TOI-199\,b: \citet{Hobson2023}; Kepler-9\,c: \citet{Holman2010, Borsato2019}; Kepler-30\,c: \citet{SanchisOjeda2012}; Kepler-553\,c: \citet{Dalba2024}.}
\end{deluxetable*}

For each observed transit, we extracted a segment of data centered on the approximate midpoint and spanning three transit durations. Slow photometric trends were modeled by fitting the out-of-transit data with a polynomial function of time, with a degree ($1$, $2$, or $3$) chosen to minimize the Bayesian information criterion \citep{Schwarz1978}. The light curve was normalized by dividing by the lowest-BIC model (usually linear). To remove outliers, we clipped points more than $5\sigma$ away from the local mean. Flux uncertainties were estimated from the standard deviation of the out-of-transit data, and an additional jitter term was included in the likelihood function to allow for excess variance.

All transits were inspected visually; those with incomplete coverage or unusually strong or complex trends were discarded. An initial fit to a \citet{Mandel&Agol2002} model was performed using a Nelder-Mead optimizer, holding all parameters fixed to values from the literature except for the transit midpoint, jitter, and a dilution factor. The transit-specific dilution factor is an additive correction to account for transit depth variations due to stellar variability or measurement systematics (such as blending with other sources in the image). These resulting transit times were used to perform a second detrending iteration, after which the normalized light curves were stacked and refitted, allowing the transit shape parameters to vary. Short- and long-cadence data were fitted separately with different jitter terms. Finally, to account for transit timing variations (TTVs), we re-measured individual transit times after holding all parameters fixed at the consensus values except the midpoint, jitter, and dilution terms, and used them to create stacked light curves as a function of time since midtransit.

Two objects required special treatment: Kepler-9\,c, which exhibits large transit duration variations \citep{Holman2010}, and Kepler-30\,c, whose transits are affected by starspot crossings \citep{SanchisOjeda2012}. For Kepler-9\,c, we fitted each transit with a multiplicative time-stretch factor, in addition to the usual three parameters (midpoint, jitter, and dilution factor). To construct the stacked light curves, we divided Kepler-9\,c's timestamps relative to mid-transit by the stretch factors, thereby adjusting all transits to have a common transit duration. For Kepler-30\,c, we manually masked prominent starspot crossings before carrying out the fitting procedure described above.

\begin{figure*}
\centering
\includegraphics[width=0.95\textwidth]{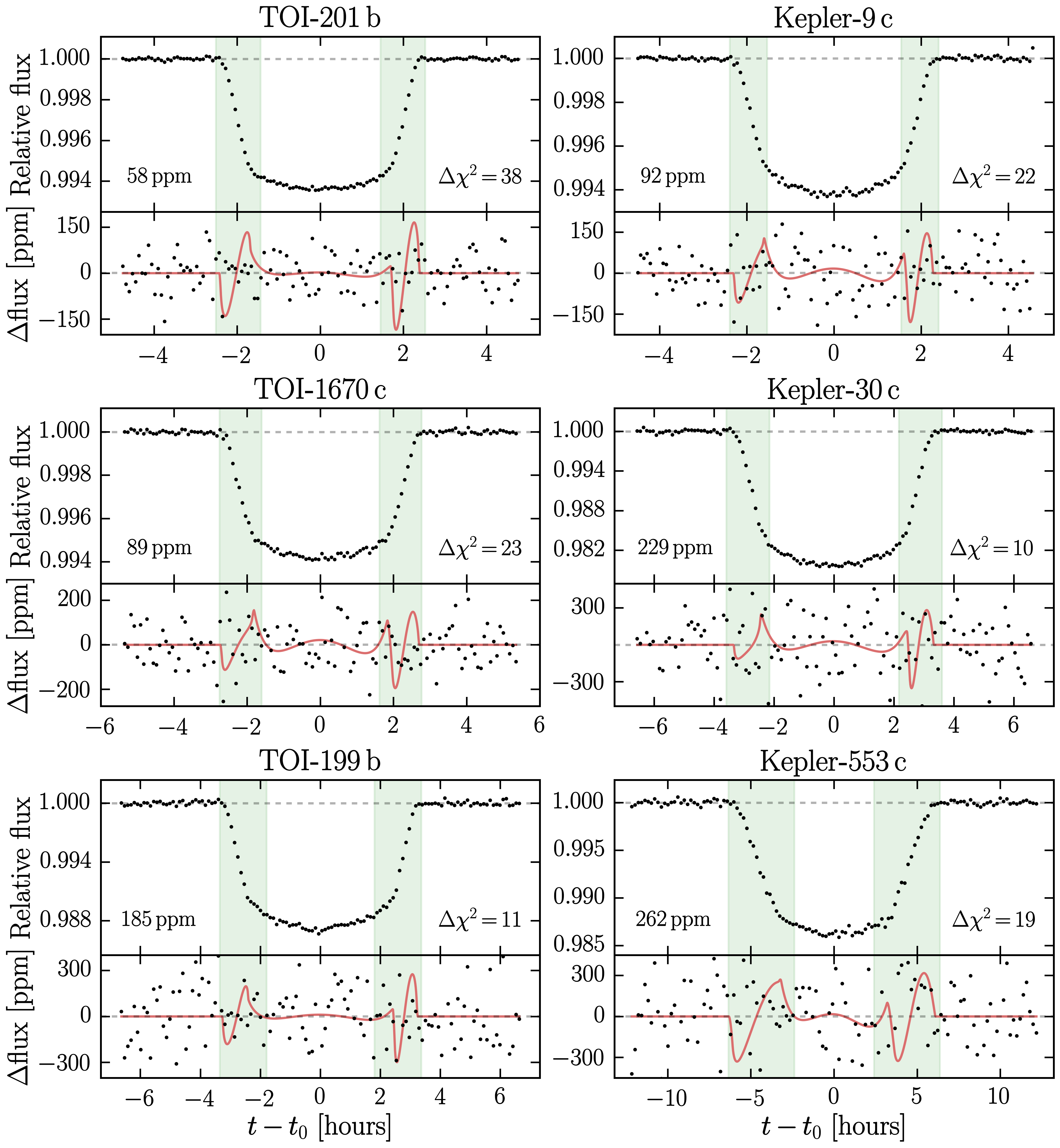}
\caption{Stacked and binned transit light curves for our six targets. In the panel beneath each light curve, black points are deviations between the data and the best-fit spherical-planet model, and red curves are deviations between a fiducial oblate-planet model with $f_\perp\,{=}\,0.2$ and $\theta_\perp\,{=}\,45^\circ$, and other parameters optimized, and the best-fit spherical-planet model. The red curves illustrate the variety of anomalies that can result from oblateness. To the left of each light curve is the residual scatter in ppm, and to the right is the $\Delta\chi^2$ by which the best-fit spherical-planet model is preferred over the fiducial oblate-planet model.}
\label{fig:light_curves}
\end{figure*}

Figure~\ref{fig:light_curves} shows the stacked light curves. For visualization purposes, the short- and long-cadence datasets were combined and binned into $125$ points. All subsequent modeling was performed on the unbinned data. We do not expect spin-precession to smear out the oblateness signal in the stacked light curves because the maximum time span of observations was $7$ years and the predicted precession periods are ${\gtrsim}\,100$\,yr, based on Equation~11 of \cite{Carter&Winn2010b} with nominal values.

\section{Oblate-planet light curve model}
\label{sec:oblate_model}

Rotational oblateness is quantified by the flattening parameter $f = 1 - \frac{b}{a}$, where $a$ is the planet's equatorial radius and $b$ is its polar radius (not to be confused with the transit impact parameter). The planet’s obliquity is the angle $\theta$ between its polar and orbital axes.

The transit light curve depends only on the shape of the planet's silhouette, assumed to be elliptical and parameterized by $f_\perp\,{=}\,1\,{-}\,\frac{b_\perp}{a_\perp}$, where $a_\perp$ and $b_\perp$ are the semimajor and semiminor axes of the projected ellipse. The planet's projected obliquity $\theta_\perp$ is the angle between the major axis of the projected ellipse and the transit chord.\footnote{Figure~1 of \citet{LammersWinn2024} illustrates the meaning of $\theta_\perp$ and $f_\perp$, although there is an error: the angle labeled $\theta_\perp$ is actually $90^\circ\,{-}\,\theta_\perp$.} Appendix~\ref{sec:HB_deriv}
provides the equations connecting intrinsic and sky-projected quantities.

We used the \texttt{squishyplanet} code to generate transit light curves for oblate planets \citep{Cassese2024}. The model has $10$ free parameters. Five are standard: mid-transit time $t_\mathrm{mid}$, scaled semimajor axis $a/R_\star$, orbital inclination $i_p$, and quadratic limb-darkening parameters $q_1$ and $q_2$ \citep{Kipping2013}. The sixth parameter is $R_\perp/R_\star$, where $R_\star$ is the stellar radius and $R_\perp$ is the planet's effective radius, defined as $a_\perp \sqrt{1\,{-}\,f_\perp}$. Two more parameters are the components $e \cos \omega$ and $e \sin \omega$ of the eccentricity vector. The remaining two parameters are the projected oblateness $f_\perp$ and projected obliquity $\theta_\perp$. We adopted uniform priors on each parameter with broad allowed ranges except for $\theta_\perp$ and $f_\perp$, which were restricted to ($-\pi$/2,\,$\pi$/2) and ($0.0$,\,$0.5$).\footnote{Although we allowed $f_\perp$ to be as large as $0.5$ for convenience, when $f\,{\gtrsim}\,0.33$ the planet's equilibrium shape is no longer accurately described as an ellipsoid \citep{Press&Teukolsky1973}.} The timescale of each transit was determined by fixing the orbital period to the literature-reported value.

To derive quantitative oblateness constraints and trace out degeneracies, we performed a Markov Chain Monte Carlo (MCMC) analysis of the stacked light curves. We employed the \texttt{emcee} code of \citet{Foreman-Mackey2013}, which implements the affine-invariant sampler of \citet{Goodman&Weare2010}. We used $400$ independent walkers, each taking $50{,}000$ steps, the first $20$\% of which were discarded as burn-in. We obtained similar results when we repeated the MCMC analysis with $150{,}000$ steps. The walkers were initialized using the best-fit spherical-planet parameters perturbed by small Gaussian random deviates. Initial values of $\theta_\perp$ and $f_\perp$ were drawn from uniform distributions spanning ($-\pi$/2,\,$\pi$/2) and ($0.0$,\,$0.5$).

The constraints on $f_\perp$ and $\theta_\perp$ are severely degenerate. It proved more insightful to examine the posteriors of $f_\perp \cos(2\theta_\perp)$ and $f_\perp \sin(2\theta_\perp)$, which isolate the even and odd components of the oblateness signal, respectively (see Figure~\ref{fig:antisymmetric} and Appendix~\ref{sec:oblate_geometry}; also used by \citealt{Liu2024}).

\section{System-by-system results}
\label{sec:analysis}

Below, we describe each target and the results of our search for oblateness-related light-curve anomalies. None of the six systems showed compelling evidence for an oblateness anomaly, but some cases turned out to be more constraining than others. For ease of comparison, we quantified the sensitivity of each dataset by calculating $\Delta \chi^2$ between the best-fit spherical-planet model and a fiducial oblate-planet model with $f_\perp\,{=}\,0.2$, $\theta_\perp\,{=}\,45^\circ$, and all other parameters optimized.

\subsection{TOI-201\,b}
\label{sec:TOI201}

TOI-201 is a bright F-type star with three known transiting companions: a warm Jupiter with a $53$~day orbit, a super-Earth on a $6$~day orbit, and a transiting brown dwarf on an $8$~year orbit \citep{Hobson2021, MaciejewskiLoboda2025, Mireles2026}. The warm Jupiter's bright host star, nearly ideal impact parameter, long expected tidal despinning timescale, and numerous available transit observations make it an outstanding target. The transit data are well-fitted by a spherical-planet model, with a residual scatter of $58$~ppm, and $\Delta \chi^2\,{=}\,38$ between the spherical-planet model and the fiducial oblate-planet model.

Figure~\ref{fig:2D_constraints} shows the posteriors of the oblateness parameters for this and the other systems. As expected, the odd component of the signal, $f_\perp \sin(2 \theta_\perp)$, is more tightly constrained than the even component, $f_\perp \cos(2 \theta_\perp)$. Although one might expect the constraints to depend only on $|\theta_\perp|$, since light curves with opposite signs of $\theta_\perp$ are related by time inversion, noise fluctuations can break the symmetry and favor one sign over another. For TOI-201\,b, the constraints were tighter for negative values than positive values. A Saturn-like oblateness can be ruled out with ${\gtrsim}\,2\sigma$ confidence over most of the range of $\theta_\perp$.

\subsection{TOI-1670\,c}
\label{sec:TOI1670}

TOI-1670 is a bright F-type star with a transiting Jupiter-sized planet on a $41$~day orbit, accompanied by a $2~R_\oplus$ planet on an $11$~day orbit. The orbit of the warm Jupiter, TOI-1670\,c, appears to be well-aligned with the host star's equator \citep{Lubin2023}. After binning {\it TESS}'s $20$ transits, and subtracting off the best-fit spherical planet model, the residual scatter is $89$~ppm.

\begin{figure*}
\centering
\includegraphics[width=0.95\textwidth]{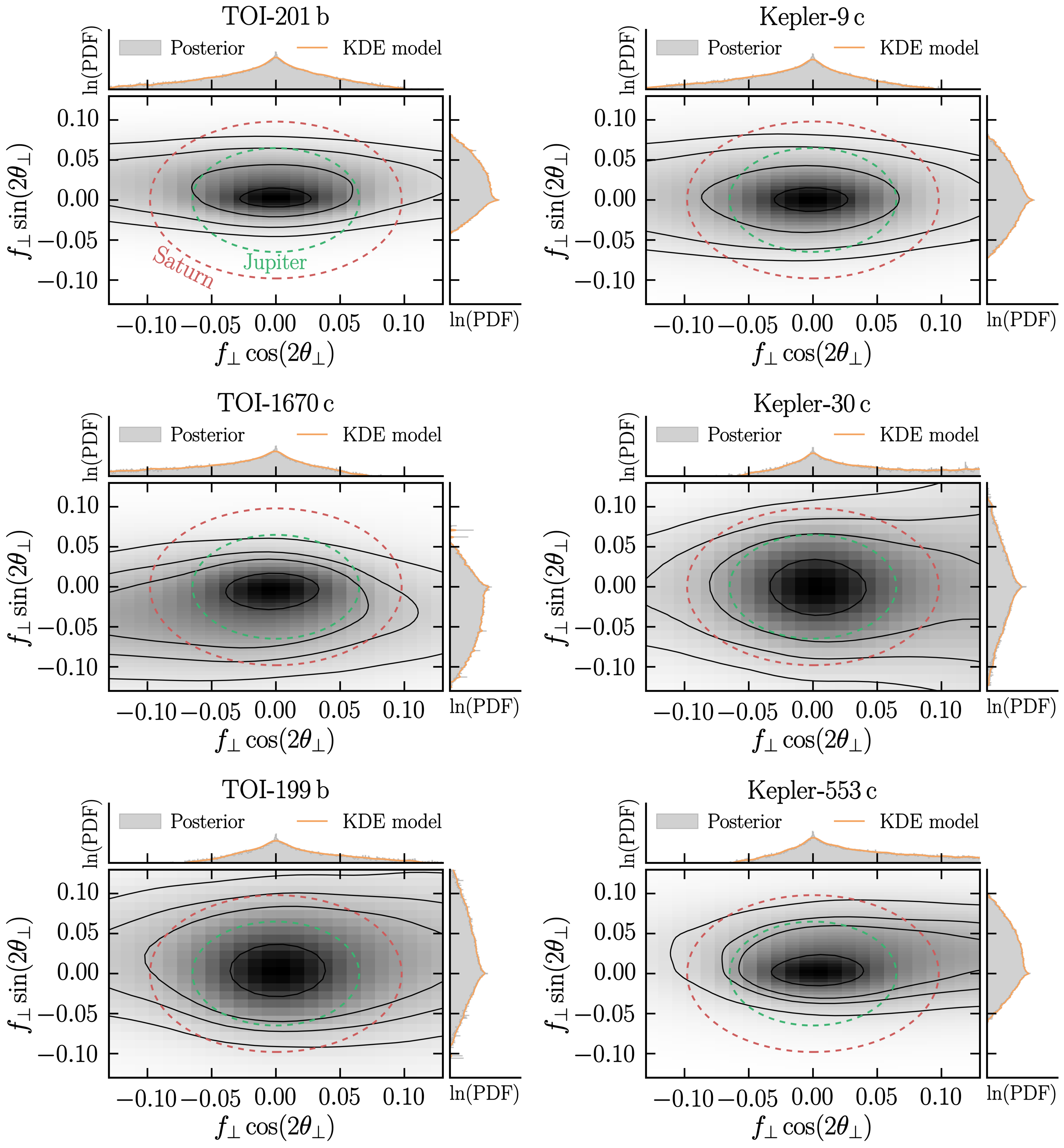}
\caption{Joint posterior probability densities of $f_\perp \cos (2\theta_\perp)$ and $f_\perp \sin (2 \theta_\perp)$. Black contours are $0.5$, $1.0$, $1.5$, and $2.0$-$\sigma$ joint confidence levels. Marginalized posteriors and their KDE approximations are shown on the top and right axes. The constraints on each planet's oblateness depend sensitively on the planet's unknown obliquity. For intermediate values of $\theta_\perp$, a Saturn-like $f_\perp$ is ruled out at ${\gtrsim}\,2\sigma$ significance for most of the systems. However, Saturn-like oblateness is allowed for other orientations. Even larger oblateness is allowed if $\theta_\perp\,{\approx}\,0^\circ$ or $90^\circ$ because the odd component of the signal vanishes.}
\label{fig:2D_constraints}
\end{figure*}

The spherical-planet model is superior to the fiducial oblate-planet model by $\Delta \chi^2\,{=}\,23$. The constraints, shown in Figure~\ref{fig:2D_constraints}, are not as tight as for TOI-201\,b. A Saturn-like projected oblateness with $\theta_\perp\,{\approx}\,45^\circ$ can be ruled out with ${>}\,2\sigma$ confidence, but for $\theta_\perp\,{\approx}\,-45^\circ$, the confidence falls to about $1.5\sigma$.

\subsection{TOI-199\,b}
\label{sec:TOI199}

TOI-199 is a bright G-type star that hosts a transiting Saturn-like planet on a $105$~day orbit. TOI-199\,b exhibits moderate TTVs (up to $1$~hour), which were combined with RV measurements to detect another giant planet on a $0.8$~AU orbit \citep{Hobson2023}.

TOI-199\,b's long, deep transit and moderate impact parameter make it a compelling oblateness target, and the planet's relatively long orbital period alleviates concerns about tidal despinning ($\tau_s\,{\approx}\,294$\,Gyr). The transit data are compatible with a spherical-planet model, with a residual scatter of $185$~ppm. It is favored over the fiducial oblate-planet model by $\Delta \chi^2\,{=}\,11$. In this case, the constraints were stronger for $\theta_\perp\,{<}\,0^\circ$. A Saturn-like $f_\perp$ can be ruled out with $2\sigma$ confidence for $\theta_\perp\,{\approx}\,-45^\circ$. For other values of $\theta_\perp$, larger-than-Saturn $f_\perp$ values are possible.

\subsection{Kepler-9\,c}
\label{sec:Kepler9}

Kepler-9 is a G-type star that hosts two giant planets nearly in $2$:$1$ mean-motion resonance, as well as a super-Earth on a $1.6$~day orbit \citep{Holman2010, Torres2011}. This was the first star known to host multiple transiting planets. Kepler-9\,c, the planet studied here, has a low bulk density of about $0.3$~g/cm$^3$. The estimated tidal despinning timescale, ${\sim}\,4$\,Gyr, is longer than our imposed minimum of $1$~Gyr but is the shortest of the planets in our sample.

The transit data are well-fitted by a spherical-planet model, leaving a residual scatter of $92$~ppm. The fiducial oblate-planet model was disfavored with $\Delta \chi^2\,{=}\,22$. The constraints were approximately symmetric about $\theta_\perp\,{=}\,0^\circ$, with a ${>}\,2\sigma$ preference against a Saturn-like $f_\perp$ for $\theta_\perp\,{\approx}\,45^\circ$ or $-45^\circ$.

\subsection{Kepler-30\,c}
\label{sec:Kepler30}

Kepler-30 is a G-type star with a system of three transiting planets. The planet studied here, Kepler-30\,c, is the largest of the trio and orbits between its neighbors, near a $2$:$1$ mean-motion resonance with the inner planet \citep{Fabrycky+2012}. It was the first {\it Kepler} system for which low mutual inclinations and spin-orbit alignment were inferred, based on an analysis of starspot-induced light-curve anomalies and variations in transit times and durations \citep{SanchisOjeda2012}.

The stacked transit light curve is well described by a spherical-planet model, giving a residual scatter of $229$~ppm. The fiducial oblate-planet model is disfavored by $\Delta \chi^2\,{=}\,10$. The posteriors imply a ${>}\,1.5\sigma$ preference against a Saturn-like $f_\perp$ for $\theta_\perp\,{\approx}\,45^\circ$ or $-45^\circ$ (Figure~\ref{fig:2D_constraints}).

\subsection{Kepler-553\,c}
\label{sec:Kepler553}

Kepler-553 is a K-type star with a warm super-Jupiter and a $5~R_\oplus$ planet on a $4$~day orbit that has been validated statistically \citep{Morton2016}. The {\it Kepler} database includes five high-quality transit observations, two in short-cadence mode and three in long-cadence mode. With its relatively wide orbit and high mass, the planet seems quite safe from tidal despinning, more so than for any other planet in our sample ($\tau_s\,{\sim}\,6\,{\times}\,10^{14}$\,yr).

The scatter about the best-fit spherical-planet model is $262$~ppm, making the light curve noisier than the rest, but the system compensates by having unusually deep transits and a nearly ideal impact parameter. The fiducial oblate-planet model is disfavored by $\Delta \chi^2\,{=}\,19$. The posteriors indicate that a Jupiter-like oblateness can be ruled out with ${>}\,2\sigma$ confidence for $\theta_\perp\,{\approx}\,-45^\circ$. As usual, for lower values of $\theta_\perp$, higher oblateness is allowed.

\section{Hierarchical Bayesian analysis}
\label{sec:BH_analysis}

Previous efforts to constrain planetary oblateness and obliquity via transit light curves have focused on upper limits derived for individual systems. Although this is a worthwhile starting point, the astrophysical implications are ambiguous because of the very strong degeneracy between projected oblateness and obliquity, and because the signal depends only on the planet's sky-projected shape. Below, we describe a method to overcome these limitations using a sample of multiple planets.

\subsection{Formalism}
\label{sec:BH_formalism}

A transit light curve dataset $\vec{d}$ provides constraints
on the parameters $\vec{x}\,{=}\,(f_\perp \cos 2 \theta_\perp,\,f_\perp \sin 2 \theta_\perp)$. The posterior probability distribution is
\begin{equation}
\label{eqn:posterior}
p(\vec{x}{|}\,\vec{d}) \propto p(\vec{d}\,{|}\,\vec{x})~p_0(\vec{x}),
\end{equation}
where $p(\vec{d}\,{|}\,\vec{x})$ is the likelihood function and $p_0(\vec{x})$ is the prior probability distribution. With a collection of datasets $\{\vec{d}_1,...,\vec{d}_N\}$ and corresponding posteriors on $\vec{x}_1, ..., \vec{x}_N$, we can constrain the hyperparameters of a model for the distribution of true oblateness $f$ and true obliquity $\theta$.

Given the limitations of the current data, we did not attempt to infer both the oblateness and obliquity distributions. Instead, we considered a few simple prescriptions for the intrinsic oblateness, informed by Jupiter and Saturn, and inferred the obliquity distribution conditional on each prescription.

We modeled the obliquities as a Fisher distribution,
\begin{equation}
\label{eqn:Fisher}
p(\theta\,{|}\,\kappa) = \frac{\kappa}{2 \sinh \kappa} \exp(\kappa \cos \theta) \sin \theta,
\end{equation}
where $\kappa$ is the concentration, the single hyperparameter \citep{Fisher1953}. The posterior for $\kappa$ is
\begin{equation}
\label{eqn:hierarchical_posterior1}
p(\kappa\,{|}\,\{\vec{d}_i\}) \propto p_0(\kappa) \prod_{i=1}^N p(\vec{d}_i\,{|}\,\kappa),
\end{equation}
where $p_0(\kappa)$ is the hyperprior on $\kappa$ and the product sequence is the hierarchical likelihood. The likelihood for an individual system is
\begin{align}
 p(\vec{d}_i\,{|}\,\kappa) &= \int\,p(\vec{d}_i\,{|}\,\vec{x}_i)\,p(\vec{x}_i\,{|}\,\kappa)\,d^2x_i\\
&\propto\int\,p(\vec{x}_i\,{|}\,\vec{d}_i)\,\frac{p(\vec{x}_i\,{|}\,\kappa)}{p_0(\vec{x}_i)}\,d^2x_i.
\label{eqn:hierarchical_likelihood}
\end{align}
In the literature, the hyperparameter likelihood is often estimated using the $K$-samples approximation \citep{Hogg2010}, which operates on MCMC samples. We encountered difficulties with this approach because a sufficiently accurate approximation required more MCMC samples than were computationally feasible to produce. Instead, we modeled the histogram of MCMC samples from each posterior $p(\vec{x}_i{|}\,\vec{d}_i)$ using a fitting function, and used those functions to compute the hierarchical posterior
\begin{equation}
\label{eqn:hierarchical_posterior2}
p(\kappa\,{|}\,\{\vec{d}_i\}) \propto p_0(\kappa) \prod_{i=1}^N \int\,\mathcal{P}_i(\vec{x}_i\,{|}\,\vec{y}_i)\,\frac{p(\vec{x}_i\,{|}\,\kappa)}{p_0(\vec{x}_i)}\,d^2x_i.
\end{equation}
Here, $\mathcal{P}_i$ is the fitting function for the posterior of system $i$, and $\vec{y}_i$ are the parameters of that function. $p_0(\vec{x}_i)\,{\propto}\,1/f_\perp$ is the prior on $\vec{x}_i$ implied by the uniform priors on $f_\perp$ and $\theta_\perp$. For the fitting functions, we used Gaussian kernel density estimation (KDE), with a bandwidth set to $2$\% of the posterior scale.\footnote{The results were insensitive to reasonable changes to the KDE bandwidth.} The remaining task was to calculate $p(\vec{x}_i\,{|}\,\kappa)$. This was accomplished analytically; see Equation~\ref{eqn:pushfoward2} in Appendix~\ref{sec:HB_deriv}.

For the hyperprior $p_0(\kappa)$, we adopted the function suggested by \citet{Fabrycky&Winn2009},
\begin{equation}
\label{eqn:Fisher_prior}
p_0(\kappa) \propto (1 + \kappa^2)^{-3/4},
\end{equation}
because it has sensible limiting behaviors: uniform in $\kappa$ for $\kappa\,{\to}\,0$, and tending toward $\kappa^{-3/2}$ as $\kappa\,{\to}\,\infty$, allowing the integral over $\kappa$ to remain finite.

\begin{figure}
\centering
\includegraphics[width=0.475\textwidth]{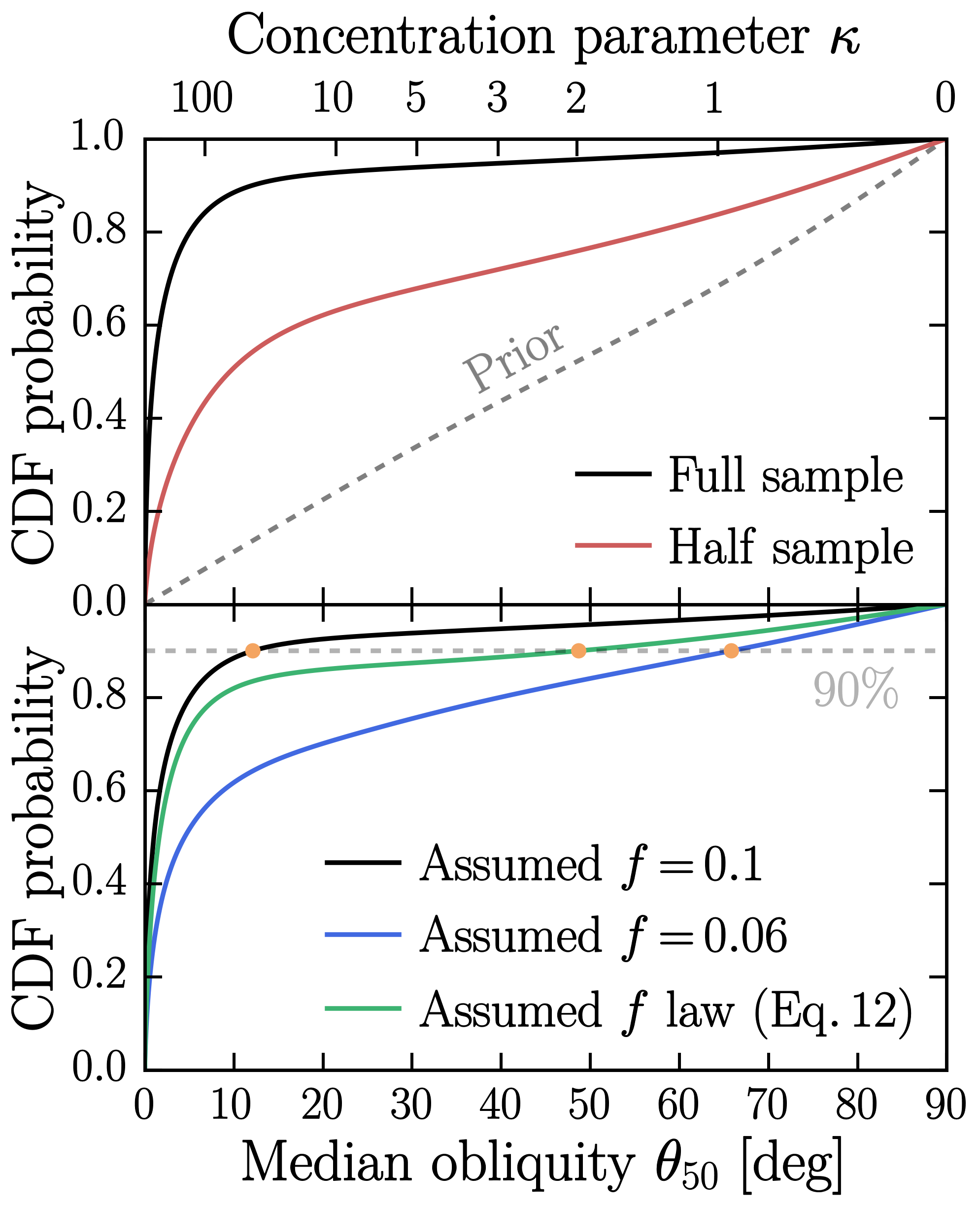}
\caption{CDF of the median obliquity, $\theta_{50}$, from our hierarchical Bayesian analysis. The upper horizontal axis translates $\theta_{50}$ into the concentration $\kappa$ of the Fisher distribution. Black curves are posteriors based on the full sample, assuming all planets have $f\,{=}\,0.1$, comparable to Saturn. The blue curve is the posterior assuming $f\,{=}\,0.06$, comparable to Jupiter. The red curve is based on half the planets; weaker constraints are obtained, reflecting the fact that useful information is distributed across all systems. The green curve shows the posterior assuming $f$ is assigned via Equation~\ref{eqn:oblateness_law}. Assuming $f\,{=}\,0.1$ or $f\,{=}\,0.06$, the data imply $\theta_{50}\,{<}\,12^\circ$ or $\theta_{50}\,{<}\,66^\circ$, respectively, at $90$\% confidence.}
\label{fig:hierarchical_CDF}
\end{figure}

A more intuitive way to parameterize the obliquity distribution is via the median
obliquity, $\theta_{50}$. For a Fisher distribution,
\begin{equation}
\label{eqn:Fisher_median}
\theta_{50} = \arccos\left(\frac{\ln\cosh \kappa}{\kappa}\right).
\end{equation}
The $\kappa$ posterior can be converted to a $\theta_{50}$ posterior using the determinant of the coordinate transformation Jacobian,
\begin{equation}
\label{eqn:theta50_conversion}
p(\theta_{50}\,{|}\,\{\vec{d}_i\}) = p(\kappa\,{|}\,\{\vec{d}_i\}) \frac{\kappa \sqrt{\kappa^2 - \ln^2 \cosh \kappa}}{\kappa \tanh \kappa - \ln \cosh \kappa}.
\end{equation}
The dashed curve in Figure~\ref{fig:hierarchical_CDF} shows the CDF of the prior for $\theta_{50}$, obtained by applying this transformation to Equation~\ref{eqn:Fisher_prior}. This curve is very close to linear, implying a nearly uniform prior on $\theta_{50}$.

\subsection{Results}
\label{sec:results}

We used the procedure described above to calculate the posterior $p(\theta_{50}\,{|}\,\{\vec{d}_i\})$ over the range $\theta_{50}\,{\in}\,(1^\circ,\,90^\circ)$. For $\theta_{50}\,{\lesssim}\,1^\circ$ ($\kappa\,{\gtrsim}\,4500$), the calculation becomes numerically challenging because $p(\vec{x}\,{|}\,\kappa)$ grows sharply (Equation~\ref{eqn:pushfoward2}), making a dense integration grid necessary. Fortunately, an analytic approximation is available in the same limit. Over the range $\theta_{50}\,{\in}\,(1^\circ,\,5^\circ)$, the posteriors are well approximated by the function $A\exp(-\beta \theta_{50}^\gamma)$, with three free parameters $A$, $\beta$, and $\gamma$. We used this analytic approximation to extrapolate the posteriors to $\theta_{50}\,{\approx}\,0^\circ$, after which we calculated cumulative distribution functions (CDFs) for $\theta_{50}$. The results are insensitive to the details of this extrapolation.

Figure~\ref{fig:hierarchical_CDF} shows posterior CDFs for the median obliquity. The black curves are posteriors conditioned on the assumption that each planet has a Saturn-like intrinsic oblateness of $f\,{=}\,0.1$. In this case, the six non-detections imply that $\theta_{50}\,{<}\,12^\circ$ with $90$\% confidence, and $\theta_{50}\,{<}\,27^\circ$ (Saturn's obliquity) with $93$\% confidence. The result is weaker when any subset of a few planets is dropped from the analysis, confirming that all members of the sample contribute useful information. Without Kepler-9\,c, which is at greatest risk of tidal despinning, $\theta_{50}\,{<}\,27^\circ$ at $87$\% confidence. The red curve shows the posterior considering only the three most informative targets, TOI-201\,b, Kepler-9\,c, and Kepler-553\,c, which is significantly broader.

The strength of the constraints depends on the assumed oblateness distribution. We performed two additional calculations based on different assumptions. The blue curve in Figure~\ref{fig:hierarchical_CDF} shows the constraints when the planets are assumed to possess a Jupiter-like oblateness of $f\,{=}\,0.06$. In this case, the $90$\% confidence limit on $\theta_{50}$ is ${<}\,66^\circ$. We also tried assigning different intrinsic oblatenesses to each planet using the rule
\begin{equation}
    \label{eqn:oblateness_law}
    f = 0.06 \left(\frac{R_p}{\mathrm{R_{Jup}}}\right)^{\!\!3} \left(\frac{\mathrm{M_{Jup}}}{M_p}\right),
\end{equation}
inspired by the leading-order scaling relation $f\,{\propto}\,\Omega^2/\rho$ \citep{dePater&Lissauer2015} and the assumption that $\Omega$ depends weakly on $R_p$ and $M_p$ \citep{Hughes2003, Scholz2018}. Equation~\ref{eqn:oblateness_law} was normalized based on Jupiter and correctly predicts $f\,{\approx}\,0.1$ for Saturn. The green curve in Figure~\ref{fig:hierarchical_CDF} shows the resulting posterior for $\theta_{50}$. In this case, the $90$\% confidence limit on $\theta_{50}$ is ${<}\,49^\circ$.

The constraints presented above are subject to the assumption that $\theta$ obeys a Fisher distribution, which is peaked near $0^\circ$ as $\kappa\,{\to}\,\infty$. Distributions peaked near $90^\circ$ or $180^\circ$ would also be consistent with the data, because the detectable component of the oblateness signal vanishes as $\theta\,{\to}\,90^\circ$ and the likelihood function is unaffected by the transformation $\theta\,{\to}\,\theta\,{+}\,180^\circ$. We expect the limits on $|\theta\,{-}\,\theta_\mathrm{center}|$ for $\theta_\mathrm{center}\,{=}\,90^\circ$ and $180^\circ$ to be similar to those derived with $\theta_\mathrm{center}\,{\approx}\,0^\circ$ (e.g., ${<}\,12^\circ$ for $f\,{\approx}\,0.1$).

\subsection{Future outlook}
\label{sec:future}

Based on the current sample, the constraints on the obliquity distribution are physically meaningful, but modest and conditional on the assumed oblateness distribution. As the sample of objects grows and detections are achieved, the data should be able to achieve tighter constraints.

A promising way forward is to conduct {\it JWST} observations of the transits of the most observationally favorable systems. Because of the telescope's larger collecting area and superior stability, {\it JWST} should be more sensitive to oblateness than {\it Kepler} or {\it TESS}. Furthermore, {\it JWST} could observe the most favorable known targets, whereas our analysis was restricted by the available archival data. To gauge the prospects, we applied the hierarchical Bayesian machinery to forecast the constraints on the obliquity distribution that might be obtained from a multi-object {\it JWST} program.

\citet{Wang&Winn2026} have predicted that ${\sim}\,5$\,--$10$ known warm Jupiters are sufficiently observationally favorable to detect Jupiter-like oblateness. Motivated by this prediction, we assumed that six high-precision oblateness constraints have been collected. For our forecast, we assumed $f\,{=}\,0.1$ for each planet and sampled obliquities $\theta$ from a Fisher distribution. We then chose six random azimuthal angles, and calculated the corresponding projected quantities $\theta_\perp$ and $f_\perp$. Lastly, we generated simulated Gaussian posteriors for $f_\perp \sin 2 \theta_\perp$ and $f_\perp \cos 2 \theta_\perp$ with means set at the true values and standard deviations of $0.005$ and $0.025$, respectively. The assumed uncertainties were based on the precision predicted by the {\it JWST} Exposure Time Calculator for a single-transit NIRSpec/G395H observation of TOI-201\,b; they are approximately five times smaller than those based on the stacked {\it TESS} data (Section~\ref{sec:analysis}). The synthetic posteriors were treated in the same way as the real posteriors in Section~\ref{sec:results}. That is, we fit the posteriors with KDE models, calculated the posterior $p(\theta_{50}\,{|}\,\{\vec{d}_i\})$, extrapolated down to $\theta_{50}\,{\approx}\,0^\circ$, and calculated CDFs.

\begin{figure}
\centering
\includegraphics[width=0.475\textwidth]{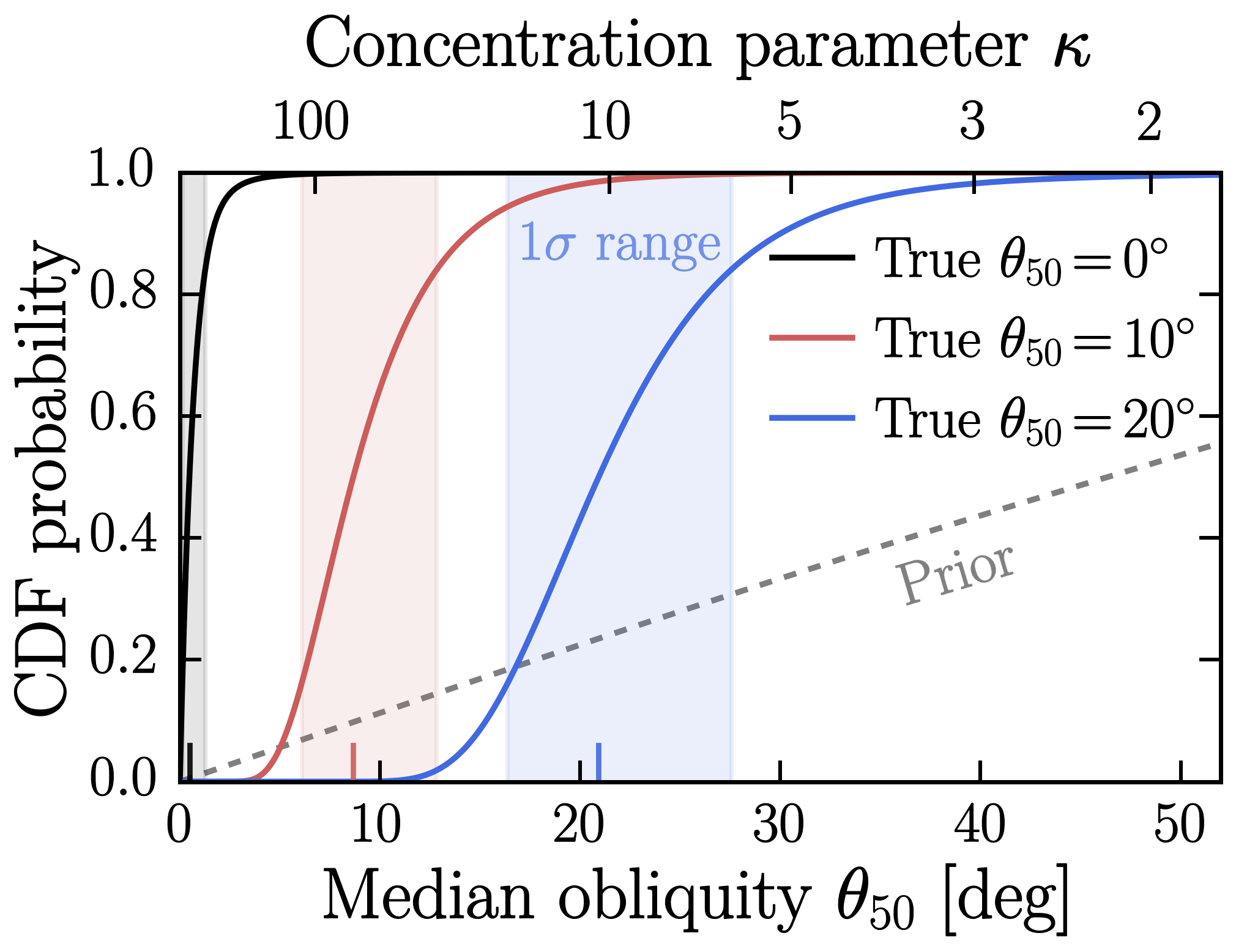}
\caption{Forecast for the constraints on the obliquity distribution provided by a sample of six objects with {\it JWST}-quality oblateness constraints (see Section~\ref{sec:future}). The three curves are CDFs for the median obliquity based on six simulated datasets created using Fisher distributions with different $\theta_{50}$ values, and assuming $f\,{=}\,0.1$. One-sigma ranges are shown as shaded regions, and median values are indicated on the horizontal axis. Six precise transit light curves of targets as observationally favorable as TOI-201\,b would provide enough information to distinguish between $\theta_{50}\,{\approx}\,0^\circ$, ${\approx}\,10^\circ$ and $\theta_{50}\,{\approx}\,20^\circ$.}
\label{fig:hierarchical_PDF_forecast}
\end{figure}

Figure~\ref{fig:hierarchical_PDF_forecast} shows the posterior CDFs for three experiments in which the obliquities were drawn from a Fisher distribution with $\theta_{50}\,{=}\,0^\circ$, $10^\circ$, and $20^\circ$. Precise oblateness constraints for six systems provide enough information to distinguish between these three scenarios --- the three CDFs in Figure~\ref{fig:hierarchical_PDF_forecast} differ significantly. The differences were weaker, but present nonetheless, when we repeated this experiment assuming Jupiter-like intrinsic oblatenesses ($f\,{=}\,0.06$).

\section{Discussion}
\label{sec:discussion}

Below, we compare our results with previous work, provide suggestions for future oblateness analyses, and conclude with a summary.

\subsection{Comparison with previous work}
\label{sec:comparison}

The six objects presented in this study led to tighter constraints than the objects studied by \citet{Zhu2014}. Our six constraints are comparable to those obtained for Kepler-51\,d using {\it JWST} observations \citep{LammersWinn2024, Liu2024}, but our analysis has the advantage of overcoming the sky-projection problem. \citet{Cassese2026} presented {\it JWST}-based results for Kepler-167\,e, but it is difficult to compare with them because they only reported $1D$ $f_\perp$ posteriors, assuming obliquities are drawn from an isotropic distribution.

\citet{Poon2025} performed a related population-level inference of the obliquity distribution of substellar companions with much wider orbits (${\gtrsim}\,50$\,AU) and greater masses (${\gtrsim}\,10\,\mathrm{M_{Jup}}$). Their analysis relied on comparing the planets' spin-axis inclinations (based on $R_p$, $v \sin i$, and $P_\mathrm{rot}$ measurements) and orbital inclinations (from astrometry). Based on the four objects for which this information is available, they reported evidence in favor of obliquities drawn from an isotropic distribution ($\kappa\,{=}\,0$) rather than Solar System-like obliquities ($\kappa\,{=}\,5$). In contrast, our results disfavor an isotropic distribution, assuming a Saturn-like oblateness. Although both results are based on relatively small samples and different modeling assumptions, the comparison hints that warm Jupiters tend to have lower obliquities than cold super-Jupiters.

\subsection{Suggestions for oblateness analyses}
\label{sec:suggestions}

Through this work, we learned a few lessons about constraining planetary oblateness/obliquity using transit photometry that might be useful to others pursuing this goal:
\begin{itemize}

    \item Careful target selection is crucial. The impact parameter is a major factor, often outweighing many other considerations. For a target with $b\,{=}\,0.1$ to be as favorable as a similar target with $b\,{=}\,0.7$, the host star must be ${\approx}\,3.5$~mag brighter.

    \item The parameters $f_\perp \sin 2 \theta_\perp $ and $f_\perp \cos 2 \theta_\perp$ are convenient because they isolate the odd- and even-symmetry components of the oblateness signal (Figure~\ref{fig:antisymmetric}; Appendix~\ref{sec:oblate_geometry}). The joint posterior of these parameters encodes the constraints on $f_\perp$ as a function of $\theta_\perp$. Reporting a single upper limit of $f_\perp$ or $f$ is not very meaningful, due to the strong dependence on the unknown angle $\theta_\perp$.

    \item One must allow $\theta_\perp$ to range from ($-90^\circ$,\,$90^\circ$), instead of the restricted range ($0^\circ$,\,$90^\circ$) adopted in some earlier works. Noise fluctuations during ingress and egress often break the expected symmetry between constraints on $\theta_\perp$ and $-\theta_\perp$.
    
\end{itemize}

\subsection{Summary}
\label{sec:summary}

With the photometric precision of today's space telescopes, transit light curves provide a means to constrain the oblateness of exoplanets, potentially providing important information about their rotational properties. We identified six warm Jupiters for which {\it Kepler} or {\it TESS} data could be used to derive meaningful constraints on oblateness and obliquity. We performed hierarchical Bayesian modeling to infer the obliquity distribution for different assumptions about oblateness. Assuming warm Jupiters are as oblate as Saturn, the data imply their median obliquity is smaller than $12^\circ$ with $90$\% confidence and smaller than Saturn's obliquity with $93$\% confidence. Alternatively, the warm Jupiters may be less oblate than Saturn. Forming with low oblateness would be somewhat surprising, because rapid rotation is a fundamental prediction of the core accretion theory. Tidal despinning would require efficient dissipation ($Q_p'\,{\lesssim}\,10^4$\,--\,$10^5$ according to Equation~\ref{eqn:tau_synch}), but tidal theory is sufficiently uncertain that this seems possible. Studying the oblateness of transiting planets might prove important for understanding the influence of tides on planetary rotation.

Assuming tidal dissipation is negligible, our results disfavor the isotropic obliquities imparted by planet-planet scattering \citep{Li&Lai2020}, and reject oblique rings large enough to produce Saturn-like effective oblateness. Although the currently available data did not provide tight enough constraints to make dramatic progress on the origin of warm Jupiters, {\it JWST} has the potential to place tighter constraints and make the first detections. The upcoming {\it PLATO} \citep{Rauer2025} and {\it Earth 2.0} \citep{Ge2022} missions might also contribute by observing many transits of warm Jupiters orbiting relatively bright stars. As a result, we expect the rotational dynamics of warm Jupiters to soon come into sharper focus.

\begin{acknowledgments}

We are grateful to the {\it Kepler} and {\it TESS} teams for providing such spectacular datasets. We thank the anonymous referee for a thoughtful report, as well as Sarah Millholland, Michael Poon, Yubo Su, and Chris Wang for useful comments and discussions. C.L. acknowledges support from a Natural Sciences and Engineering Research Council of Canada (NSERC) Postgraduate Scholarship.

The data presented in this article were obtained from the Mikulski Archive for Space Telescopes (MAST) at the Space Telescope Science Institute. The observations can be accessed at \dataset[10.17909/pag7-py03]{https://doi.org/10.17909/pag7-py03}.

We are pleased to acknowledge that the work reported herein was substantially performed using the Princeton Research Computing resources at Princeton University, which is a consortium of groups led by the Princeton Institute for Computational Science and Engineering (PICSciE) and the Office of Information Technology’s Research Computing.

\end{acknowledgments}

\appendix

\section{Even/odd decomposition of the oblateness signal}
\label{sec:oblate_geometry}

Here, we separate the even and odd components of the oblateness signal and explain why the odd signal is maximized when the transit impact parameter is $b\,{=}\,2^{-1/2}\,{\approx}\,0.7$. See also Appendix~A of \citet{Wang&Winn2026}.

Let the planet's sky projection be an ellipse with semimajor axis $a_\perp$, semiminor axis $b_\perp\,{=}\,a_\perp(1\,{-}\,f_\perp)$, and projected obliquity $\theta_\perp$. The ellipse's half-width along an angle $\alpha$ is
\begin{equation}
\label{eqn:half_width}
h(\alpha) = \sqrt{a_\perp^2 \cos^2(\alpha - \theta_\perp) + b_\perp^2 \sin^2(\alpha-\theta_\perp)}.
\end{equation}
The limb normals at ingress and egress make angles $-\psi$ and $\pi\,{+}\,\psi$ with the transit chord, where $\psi\,{=}\,\arcsin b$. To first order in $f_\perp$,
\begin{align}
    h_\mathrm{ing}~{\approx}&~a_\perp \left[1 - f_\perp \sin^2(\psi + \theta_\perp)\right],\\
    h_\mathrm{egr}~{\approx}&~a_\perp \left[1 - f_\perp \sin^2(\psi - \theta_\perp)\right].
\end{align}
The ingress duration $\tau_\mathrm{ing}$ is proportional to $h_\mathrm{ing}$. Likewise, $\tau_\mathrm{egr} \propto h_\mathrm{egr}$. The ingress/egress duration of a spherical planet of the same area, $\tau_\mathrm{sph}$, is proportional to $R_\perp\,{=}\,\sqrt{a_\perp\,b_\perp}\,{\approx}\,a_\perp(1\,{-}\,f_\perp/2)$.
Thus,
\begin{align}
    \frac{\tau_\mathrm{ing}}{\tau_\mathrm{sph}} {\approx}&~ 1 + \frac{f_\perp}{2}\cos(2\theta_\perp + 2\psi),\\
    \frac{\tau_\mathrm{egr}}{\tau_\mathrm{sph}} {\approx}& ~1 + \frac{f_\perp}{2}\cos(2\theta_\perp - 2\psi),
\end{align}
from which it follows
\begin{align}
    \frac{\tau_\mathrm{ing} + \tau_\mathrm{egr}}{2\tau_\mathrm{sph}} &\approx 1 + \frac{f_\perp}{2} \cos(2 \theta_\perp) \cos(2 \psi) \label{eqn:tau_mean}\\
    \frac{\tau_\mathrm{ing} - \tau_\mathrm{egr}}{2\tau_\mathrm{sph}} &\approx -\frac{f_\perp}{2} \sin(2 \theta_\perp) \sin(2 \psi). \label{eqn:tau_diff}
\end{align}
The first of these equations sets the scale of the even component of the oblateness signal; it describes the common factor by which both ingress and egress durations are affected. The second equation relates to the odd component; it describes the breaking of the ingress/egress symmetry. The even and odd components of the signal are therefore governed by $f_\perp \cos(2\theta_\perp)$ and $f_\perp \sin(2\theta_\perp)$, respectively, and the odd component scales with $\sin(2 \psi)\,{=}\,2b\sqrt{1\,{-}\,b^2}$, which is maximized at $b\,{=}\,2^{-1/2}$. This optimal impact parameter was found earlier in numerical experiments by \cite{Barnes&Fortney2003} and \cite{Zhu2014}.

\section{Distribution of projected-oblateness parameters}
\label{sec:HB_deriv}

To calculate the hierarchical posterior in Equation~\ref{eqn:hierarchical_posterior2}, we need the distribution of $\vec{x}\,{=}\,(f_\perp \cos 2 \theta_\perp ,\,f_\perp \sin 2 \theta_\perp)$ implied by an assumed intrinsic oblateness $f$ and a Fisher obliquity distribution $p(\theta | \kappa)$. For a planet with an edge-on orbit ($i\,{=}\,90^\circ$),\footnote{This approximation is valid for our targets, which differ from $i\,{=}\,90^\circ$ by ${<}\,2^\circ$.} the projected and intrinsic quantities are related by
\begin{equation}
f_\perp = 1 - \sqrt{\sin^2\theta' + (1 - f)^2 \cos^2\theta'},~
\cos^2 \theta' = \sin^2 \theta \sin^2 \phi + \cos^2 \theta,~\mathrm{and}~
\tan \theta_\perp = \tan \theta \sin \phi,
\label{eqn:theta_perp}
\end{equation}
where $\phi$ is the azimuth of the line of nodes between the planet’s equatorial and orbital planes. Define
\begin{equation}
\label{eqn:param_definitions}
u\,{\equiv}\,\cos \theta,~s\,{\equiv}\,\sin \phi,~q\,{\equiv}\cos^2 \theta',~\mathrm{and}~B\,{\equiv}\,1\,{-}\,(1\,{-}\,f)^2.
\end{equation}
Then
\begin{align}
    f_\perp &= 1 - \sqrt{1 - B\,q}, \label{eqn:f_perp2} \\
    \tan \theta_\perp &= \frac{s\,\sqrt{1 - u^2}}{u},~\mathrm{and} \label{eqn:theta_perp2}\\
    \cos^2 \theta' &= (1 - u^2)s^2 + u^2. \label{eqn:cos_theta2}
\end{align}
It follows from the last equation that $|u|\,{=}\,\sqrt{q}\,\cos \theta_\perp$. The Fisher distribution $p(\theta\,{|}\,\kappa)$ can be written in terms of $u$:
\begin{equation}
p(u\,{|}\,\kappa) = \frac{\kappa}{2\sinh \kappa} \exp(\kappa\,u).
\end{equation}
It is physically reasonable to assume that $\phi$ is randomly distributed between (0,\,2$\pi$) because typical warm-Jupiter spin precession periods are ${\sim}\,10^5$ times shorter than system ages. For a uniform $\phi$ distribution, the joint probability $p(u,s\,{|}\,\kappa)$ is
\begin{equation}
p(u,s\,{|}\,\kappa) = \frac{\kappa}{2\pi\sinh \kappa} \frac{\exp(\kappa\,u)}{\sqrt{1 - s^2}}.
\end{equation}
The variables $(u,s)$ can be transformed into $(f_\perp,\,\theta_\perp)$ using the Jacobian determinant
\begin{equation}
\left|\frac{\partial(f_\perp,\,\theta_\perp)}{\partial(u,\,s)}\right| = \frac{B\sqrt{1 - u^2}}{1-f_\perp},
\end{equation}
which can be calculated using Equations~\ref{eqn:param_definitions}\,--\,\ref{eqn:cos_theta2}. Adding the probabilities $p(u,s\,{|}\,\kappa)$ and $p(-u,-s\,{|}\,\kappa)$ and dividing by the determinant yields
\begin{align}
p(f_\perp,\theta_\perp\,{|}\,\kappa) = \frac{\kappa}{\pi\sinh \kappa} \frac{\cosh(\kappa\,u)}{\sqrt{1 - q}} \frac{(1 - f_\perp)}{B},\label{eqn:pushfoward}~\mathrm{where}\\
B = 2f - f^2,~q = \frac{2f_\perp - f_\perp^2}{B},~|u| = \sqrt{q}\cos \theta_\perp.
\end{align}
We applied the identity $\sqrt{1-s^2}\,\sqrt{1 - u^2}\,{=}\,\sqrt{1-q}$ to simplify Equation~\ref{eqn:pushfoward}. In terms of $\vec{x}\,{=}\,(f_\perp \cos 2 \theta_\perp ,\,f_\perp \sin 2 \theta_\perp)$, this becomes
\begin{equation}
\label{eqn:pushfoward2}
p(\vec{x}\,{|}\,\kappa) = \frac{\kappa}{2\pi \sinh \kappa} \frac{\cosh(\kappa\,u)}{\sqrt{1 - q}} \frac{(1 - f_\perp)}{B f_\perp}.
\end{equation}
 
\bibliography{refs}{}
\bibliographystyle{aasjournalv7}

\end{document}